\documentclass[conference]{IEEEtran}
\usepackage[utf8]{inputenc}
\usepackage{graphicx}
\usepackage{amsmath}
\usepackage{amssymb}
\usepackage{cite}
\usepackage{url}

\usepackage{booktabs}
\usepackage{algorithm}
\usepackage{algpseudocode}
\usepackage{caption}

\usepackage{caption}
\usepackage{float} 
\usepackage{stfloats}
\usepackage{siunitx} 

\begin{document}

\title{The Wisdom of the Crowd: High-Fidelity Classification of Cyber-Attacks and Faults in Power Systems Using Ensemble and Machine Learning}

\author{
\IEEEauthorblockN{Emad Abukhousa, Syed Sohail Feroz Syed Afroz, Fahad Alsaeed, \\
Abdulaziz Qwbaiban, Saman Zonouz, and A.P. Sakis Meliopoulos}
\IEEEauthorblockA{\textit{School of Electrical and Computer Engineering, Georgia Institute of Technology, Atlanta, GA, USA} \\
\{emadak, safroz7, falsaeed3, aqwbaiban3, szonouz6, sakis.m\}@gatech.edu}
}
\maketitle

\begin{abstract} 
	This paper presents a high-fidelity evaluation framework for machine learning (ML)-based classification of cyber-attacks and physical faults using electromagnetic transient simulations with digital substation emulation at 4.8 kHz. Twelve ML models, including ensemble algorithms and a multi-layer perceptron (MLP), were trained on labeled time-domain measurements and evaluated in a real-time streaming environment designed for sub-cycle responsiveness. The architecture incorporates a cycle-length smoothing filter and confidence threshold to stabilize decisions. Results show that while several models achieved near-perfect offline accuracies (up to 99.9\%), only the MLP sustained robust coverage (98–99\%) under streaming, whereas ensembles preserved perfect anomaly precision but abstained frequently (10–49\% coverage). These findings demonstrate that offline accuracy alone is an unreliable indicator of field readiness and underscore the need for realistic testing and inference pipelines to ensure dependable classification in inverter-based resources (IBR)-rich networks.
\end{abstract}


\IEEEpeerreviewmaketitle
\begin{figure*}[!t]
\centering
\includegraphics[width=\textwidth]{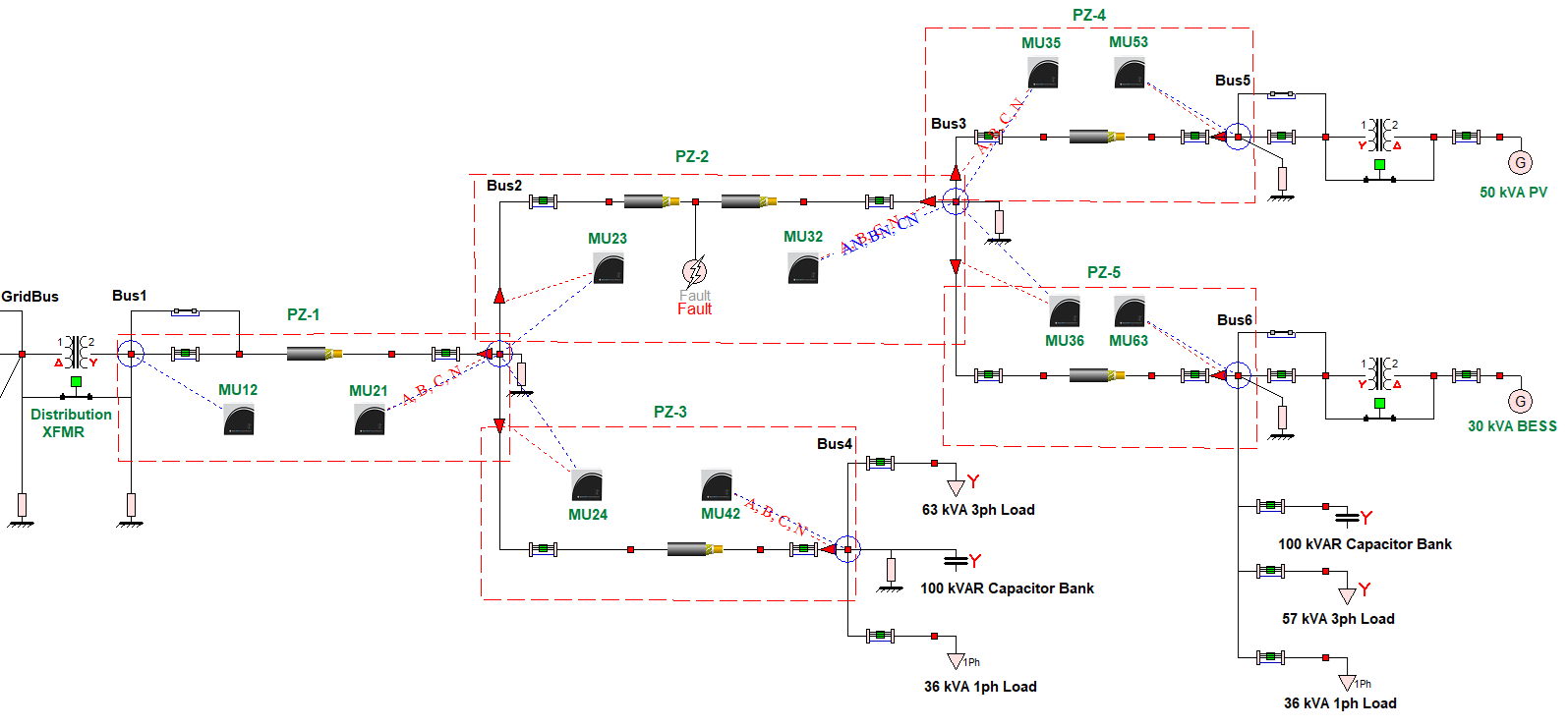}
\caption{High-fidelity microgrid testbed with inverter-based resources (IBR) integration.}
\label{fig:microgrid}
\end{figure*}
\section{Introduction}
The evolution of electric power systems toward smarter, highly automated grids has been accelerated by the proliferation of inverter-based resources (IBRs), pervasive sensing, and advanced communications. While these developments enhance flexibility, sustainability, and situational awareness, they also alter system dynamics and introduce new vulnerabilities in physical and cyber domains \cite{sridhar2012cyber, manias2024trends}. The convergence of infrastructure with cyber networks—encompassing substation automation, SCADA, and wide-area monitoring—has expanded the attack surface. Cyber intrusions such as false data injection (FDI), denial-of-service (DoS), and replay attacks can mimic or exacerbate physical disturbances, making detection and classification difficult \cite{ashok2017cyber,zonouz2012scpse}. Accurate classification of faults and cyber-attacks is critical, as attackers may inject false data or imitate faults, causing unwarranted trips or preventing necessary ones, leading to outages or equipment damage.

The growing penetration of IBRs fundamentally alters fault behavior. Unlike synchronous machines, IBRs limit fault currents—typically to 1.1–1.5 per unit of nominal output—and often generate non-sinusoidal waveforms due to semiconductor-based switching devices \cite{reno2021influence,reno2021challenges}. These characteristics degrade the effectiveness of traditional overcurrent, distance, and directional relay schemes, particularly in multi-source, bidirectional flow environments. Moreover, substation protection systems relying on IEC 61850-9-2LE sampled values are vulnerable to cyber manipulation, potentially causing misoperations \cite{manias2024trends}.

To address these challenges, the research community has explored various machine learning (ML) and deep learning (DL) techniques for anomaly detection in power systems \cite{sahani2023machine}. Approaches include deep capsule convolutional neural networks \cite{li2023identification}, ensemble-based intrusion detection systems \cite{alkasassbeh2021ensemble,al2023ensemble}, and hybrid methods that combine dynamic state estimation (DSE) with DL for fault classification in IBR-rich networks, such as the work of \cite{chencaisakis2024dl}, which used DSE plus a 1D CNN in 4.8 kHz WinIGS electromagnetic transient (EMT) simulations and achieved high accuracy for twelve fault types but did not address cyberattacks. In a related work, Alsaeed \textit{et al.} \cite{alsaeed2025anomaly} extended DSE + DL methods to include cyberattacks in high-fidelity simulations, but examined only one DL model, lacked multi-model benchmarking, and did not quantify offline-to-real-time performance shifts.

Many reported ML solutions claim high accuracy—often exceeding 95\%—but are frequently based on low-fidelity or event-simplified datasets \cite{johnson2014machine,gholami2022active}, omitting transient noise, alignment errors, and class imbalances seen in real systems. Moreover, few studies evaluate anomaly detection under live, cycle-by-cycle conditions, leaving performance under operational constraints unverified.

This paper addresses these critical gaps by presenting a comprehensive comparative study of 12 ML models for classifying both physical faults and cyber-attacks. The study is grounded in a high-fidelity, multi-class dataset from detailed WinIGS EMT simulations \cite{pserc2018t59g, winigs2025}. The research is guided by three key questions: (1) How does model performance on high-fidelity data compare with claims based on lower-fidelity datasets? (2) Which evaluation metrics best reveal the impact of class imbalance and the challenges of real-time anomaly detection? and (3) In the context of real-time anomaly detection for critical infrastructure, which "crowd" is more effective—the diversity of ensemble methods (more trees) or the adaptive learning of neural networks (more neurons)?

The contributions of this work are threefold. First, we introduce a high-fidelity evaluation framework using COMTRADE-format data from a realistic digital substation simulation, enabling robust and transparent performance assessment. Second, we compare 12 ML and DL algorithms in both static (offline) and dynamic (streaming) scenarios. Third, we enhance real-time performance via moving average filtering and confidence thresholding for more reliable decision-making. Finally, to foster reproducibility and further research, the dataset and implementation code have been open-sourced and are publicly available~\cite{T4Tech}, helping to bridge the gap between laboratory-tested models and field-deployable protection systems.

\section{Methodology}
The methodology for this study was designed as a comprehensive, multi-phase process to rigorously develop, evaluate, and validate machine learning models for anomaly detection and classification in a high-fidelity power system environment.

\subsection{Testbed and Dataset Generation}

A high-fidelity power system testbed, simulated in WinIGS, is employed in this study. The model, depicted in Fig.~\ref{fig:microgrid}, captures the dynamic interactions between synchronous generators and IBRs under realistic operating conditions. Key components include a 50 kVA photovoltaic (PV) system, a 30 kVA battery energy storage system (BESS), step-down transformers (115/13.8 kV, 34.5/13.8 kV), distribution lines, and a variety of loads, including three-phase (63 kVA, 57 kVA) and single-phase (two at 36 kVA) loads, as well as capacitor banks. The presence of both large synchronous generation and IBRs, integrated with protective relaying and digital substation elements, creates a representative environment for evaluating protection and anomaly detection in modern grids. Although the system exhibits the complexity of a utility-connected network, its configuration corresponds to a high-fidelity microgrid, making it particularly valuable for studying the unique operational and protection challenges posed by high penetration of distributed resources.

The data generation process follows the workflow of a modern digital substation compliant with the IEC 61850 standard. Current and voltage signals were acquired by virtual Merging Units (MUs), specifically MU32 and MU23, which function as the digital interface for protective relays. The data was sampled at a rate of 4.8 kHz, which corresponds to 80 samples per cycle for a 60 Hz system. This high sampling rate, with a time step ($\Delta$t) of approximately 208 $\mu$s, enables the capture of fast transient dynamics and harmonic content characteristic of IBRs and certain cyber-attacks. The final dataset was stored in COMTRADE (Common Format for Transient Data Exchange) format, the industry standard for transient data recording.

Two datasets were generated: one for model training and one for real-time prediction. The training dataset spans 22.00 seconds and contains 18 distinct classes, comprising one normal operation class and 17 classes representing various physical faults and cyber-attacks. Anomalies were injected at controlled intervals, with normal operation periods interspersed to mimic real-world conditions. The prediction dataset is a separate 6-second sequence containing 6 anomalies. Both datasets include 14 features, consisting of three-phase voltages and currents from MU32 and MU23. Table~\ref{tab:anomaly_classes} lists the anomaly classes included in the training dataset.

\begin{table}[!t]
\caption{Anomaly Classes in the Training Dataset}
\label{tab:anomaly_classes}
\centering
\begin{tabular}{@{}llc@{}}
\toprule
Class ID & Description & Event Time (s) \\
\midrule
0 & Normal Operation & Multiple \\
1 & Single Line Fault A-N & 1.00--1.50 \\
2 & Single Line Fault B-N & 2.00--2.50 \\
3 & Single Line Fault C-N & 3.00--3.50 \\
4 & CT Ratio Attack on MU32 & 4.00--4.50 \\
5 & Double Line Fault A-B & 6.00--6.50 \\
6 & Double Line Fault A-C & 7.00--7.50 \\
7 & Double Line Fault B-C & 8.00--8.50 \\
8 & CT Ratio Attack on MU23 & 9.00--9.50 \\
9 & DLG Fault AB-N & 11.00--11.50 \\
10 & DLG Fault AC-N & 12.00--12.50 \\
11 & DLG Fault BC-N & 13.00--13.50 \\
12 & PT Ratio Attack on MU32 & 14.00--14.50 \\
13 & PT Ratio Attack on MU23 & 15.00--15.50 \\
14 & 3 Lines Fault AB-C & 17.00--17.50 \\
15 & 3 Lines Fault ABC-N & 18.00--18.50 \\
16 & GPS Spoofing on MU32 & 20.00--20.50 \\
17 & GPS Spoofing on MU23 & 21.00--21.50 \\
\bottomrule
\end{tabular}
\end{table}

\subsection{Machine Learning Framework}

A rigorous machine learning framework was developed to support both offline training (Phase~1) and real-time prediction (Phase~2) evaluations.

\subsubsection{Data Preparation and Preprocessing}

The raw COMTRADE data underwent several preprocessing steps. The dataset was first inspected for missing values, which were removed. The timestamp feature was excluded to prevent potential data leakage. The remaining data were separated into a feature matrix $X$ and a target vector $y$. The 18 anomaly classes were label-encoded into a contiguous integer range. The dataset was then partitioned into 80\% training and 20\% testing subsets using stratified sampling. All features were standardized to have zero mean and unit variance according to
\begin{equation}
x_{\mathrm{scaled}} = \frac{x - \mu}{\sigma}
\end{equation}
where $\mu$ and $\sigma$ represent the mean and standard deviation, respectively, computed from the training set.
\\
\subsubsection{Model Selection and Training}

Twelve machine learning algorithms were evaluated: AdaBoost, Decision Tree, Extra Trees, Gradient Boosting, k-Nearest Neighbors (k-NN), Logistic Regression, Na\"{i}ve Bayes, Random Forest, Support Vector Machine (SVM) with a radial basis function kernel, and three Multi-Layer Perceptron (MLP) neural network configurations. Each model was trained using the standardized training dataset. For algorithms that supported it, class weights were adjusted to mitigate imbalance. Hyperparameter tuning was applied to selected algorithms via grid search with three-fold cross-validation.

\subsubsection{Phase~2: Real-Time Prediction and Post-Processing}
In Phase~2, the pre-trained models were evaluated on an unseen $6$-second event dataset containing both normal and anomalous conditions.  
Ground-truth labels $y_{\mathrm{true}}(i)$ were generated programmatically by mapping each sample time $t_i$ to known anomaly intervals:
\begin{equation}
y_{\mathrm{true}}(i) =
\begin{cases}
L_m, & t_i \in [a_m, b_m] \ \text{for some event } m,\\
0, & \text{otherwise (Normal)},
\end{cases}
\label{eq:groundtruth}
\end{equation}
where $L_m$ is the class label for event $m$, and $(a_m,b_m)$ are its start and end times.

The data stream was sampled at $f_s=\SI{4.8}{kHz}$, with the nominal grid frequency $f_{\mathrm{grid}}=\SI{60}{Hz}$, giving the number of samples per cycle as
\begin{equation}
N_{\mathrm{cyc}} = \frac{f_s}{f_{\mathrm{grid}}} = \frac{4800}{60} = 80.
\label{eq:Ncyc}
\end{equation}

At each sample $i$, the model outputs a vector of class probabilities $P(i,k)$.  
To suppress high-frequency noise and misclassification jitter, a one-cycle centered moving average with nearest-edge padding is applied:
\begin{equation}
P_{\text{avg}}[i,k] = \frac{1}{N_{\mathrm{cyc}}} \sum_{j=i-N_{\mathrm{half}-1}}^{i+N_{\mathrm{half}}} P(j,k),
\label{eq:movavg}
\end{equation}
where $N_{\mathrm{half}} = \lfloor N_{\mathrm{cyc}}/2 \rfloor$ is the half-window size.

The \textbf{model confidence level} is then defined as
\begin{equation}
c(i) = \max_{k} \ P_{\text{avg}}[i,k],
\label{eq:confidence}
\end{equation}
with predicted class
\[
\hat{k}(i) = \arg\max_{k} \ P_{\text{avg}}[i,k].
\]
A final decision is made according to the rule:
\begin{equation}
y(i) =
\begin{cases}
\hat{k}(i), & c(i) \ge \tau,\\
-1, & \text{(abstain)},
\end{cases}
\label{eq:decision}
\end{equation}
where $\tau$ is the confidence threshold. Abstentions prevent low-confidence, potentially incorrect outputs.  
This process introduces a fixed look-ahead of $N_{\mathrm{half}}$ samples ($\approx\SI{8.33}{ms}$ for $N_{\mathrm{cyc}}=80$) and achieves sub-cycle responsiveness. In this study, we use $\tau=0.6$.

The complete streaming inference and decision-layer logic is summarized in Algorithm~\ref{alg:streaming}.

\begin{algorithm}
\caption{Streaming Inference with Decision Layer}
\label{alg:streaming}
\begin{algorithmic}[1]
\State Initialize ring buffer $B$ of length $N_{\mathrm{cyc}}$
\For{each new sample $i$}
    \State $p \gets \text{model.predict\_proba}(x[i])$
    \State $B.\text{push}(p)$
    \If{$|B| = N_{\mathrm{cyc}}$}
        \State $q \gets \text{mean}(B)$ \Comment{centered window of 80 samples}
        \State $\text{conf} \gets \max(q)$; $\text{cls} \gets \arg\max(q)$
        \If{$\text{conf} \ge \tau$}
            \State $\text{emit}(\text{index } i - N_{\mathrm{half}},\ \text{cls})$
        \Else
            \State $\text{emit}(\text{index } i - N_{\mathrm{half}},\ -1)$
        \EndIf
        \State $B.\text{pop\_left}()$ \Comment{slide the window}
    \EndIf
\EndFor
\end{algorithmic}
\end{algorithm}

\begin{table*}[!h]
\caption{Comparison of Phase 1 Offline Evaluation and Phase 2 Streaming Evaluation}
\label{tab:phase_comparison}
\centering
\begin{tabular}{@{}lcccc|ccc@{}}
\toprule
 & \multicolumn{4}{c|}{\textbf{Phase 1: Offline Evaluation}} & \multicolumn{3}{c}{\textbf{Phase 2: Streaming Evaluation}} \\
\cmidrule(lr){2-5} \cmidrule(lr){6-8}
\textbf{Model} & Accuracy & Precision & Recall & F1-Score & Overall Acc. & Anomaly Acc. & Coverage (\%) \\ \midrule
Random Forest & 0.9988 & 0.9988 & 0.9988 & 0.9988 & 0.9985 & 1.0000 & 48.7 \\
MLP (3 Hidden, Wider) & 0.9988 & 0.9988 & 0.9988 & 0.9988 & 0.9986 & 0.9969 & 98.8 \\
SVM (RBF) & 0.9986 & 0.9986 & 0.9986 & 0.9986 & 0.9895 & 0.9298 & 95.9 \\
MLP (3 Hidden) & 0.9986 & 0.9986 & 0.9986 & 0.9986 & 0.9935 & 0.9657 & 99.0 \\
Gradient Boosting & 0.9985 & 0.9985 & 0.9985 & 0.9985 & 0.9983 & 0.9957 & 95.7 \\
MLP (2 Hidden) & 0.9978 & 0.9978 & 0.9978 & 0.9978 & 0.9926 & 0.9584 & 98.0 \\
K-Nearest Neighbors & 0.9977 & 0.9977 & 0.9977 & 0.9977 & 0.9911 & 0.9424 & 96.1 \\
Extra Trees & 0.9969 & 0.9969 & 0.9969 & 0.9969 & 1.0000 & 1.0000 & 10.8 \\
Decision Tree & 0.8440 & 0.8887 & 0.8440 & 0.8290 & 0.8797 & 0.8845 & 14.7 \\
Naive Bayes & 0.7354 & 0.7323 & 0.7353 & 0.7257 & 0.9909 & 1.0000 & 9.9 \\
Logistic Regression & 0.1847 & 0.1704 & 0.1846 & 0.1592 & 0.0000 & 0.0000 & 0.0 \\
AdaBoost & 0.1741 & 0.1481 & 0.1743 & 0.1217 & 0.0000 & 0.0000 & 0.0 \\ \bottomrule
\end{tabular}
\end{table*}

\section{Results}
The performance of the twelve models was evaluated in two phases: offline training and validation (Phase 1) and a real-time streaming simulation (Phase 2).

\begin{figure*}[!b]
\centering
\includegraphics[width=\textwidth]{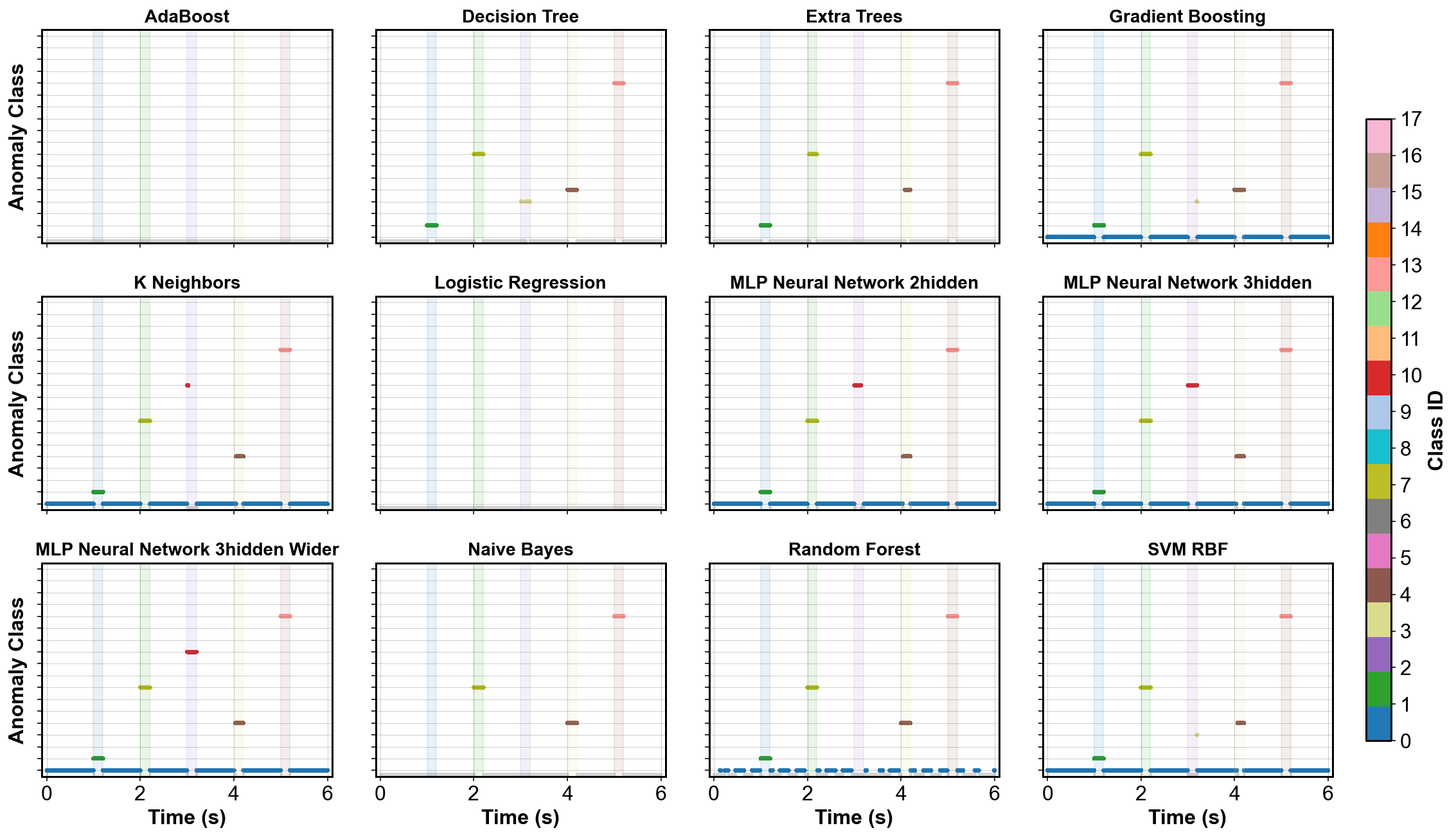}
\caption{Model responses to five sequential anomalies: SLG A-N (1.0--1.2 s), LL B-C (2.0--2.2 s), DLG AC-N (3.0--3.2 s, red), CT ratio attack MU32 (4.0--4.2 s), and PT ratio attack MU23 (5.0--5.2 s), each $\sim$0.2 s duration.}
\label{fig:model_comparison}
\end{figure*}
\subsection{Phase 1: Offline Training and Evaluation}
In Phase 1, models were trained on a balanced, high-fidelity dataset. 
Random Forest and the MLP with 3 wider hidden layers were the top performers, 
both achieving an accuracy of 99.88\%. 
The overall performance metrics are summarized in Table \ref{tab:phase_comparison}.

\subsection{Phase 2: Real-Time Streaming Evaluation}
Phase 2 evaluated model performance on a 6-second streaming dataset containing five sequential anomalies. This scenario revealed substantial differences between models when operating in a real-time, sequential environment. While most models maintained high overall accuracy, metrics focusing specifically on anomaly periods and classification coverage exposed critical operational distinctions (Table~\ref{tab:phase_comparison}).

The MLP neural networks achieved the highest classification coverage (98--99\%), meaning they produced confident classifications for nearly the entire evaluation window. In contrast, ensemble methods such as Random Forest (48.7\%) and Extra Trees (10.8\%) were more conservative, generating fewer predictions but maintaining perfect accuracy during the anomaly segments they did classify. This indicates that while MLPs are more aggressive in decision-making, ensembles tend to abstain more frequently unless highly certain.

Among the anomalies, the double line-to-ground (DLG) AC-N fault occurring between 3.0--3.2~s proved to be the most challenging, causing accuracy to drop significantly for many models, with some averaging as low as 40\% during this interval.

\begin{figure}[!t]
\centering
\includegraphics[width=\columnwidth]{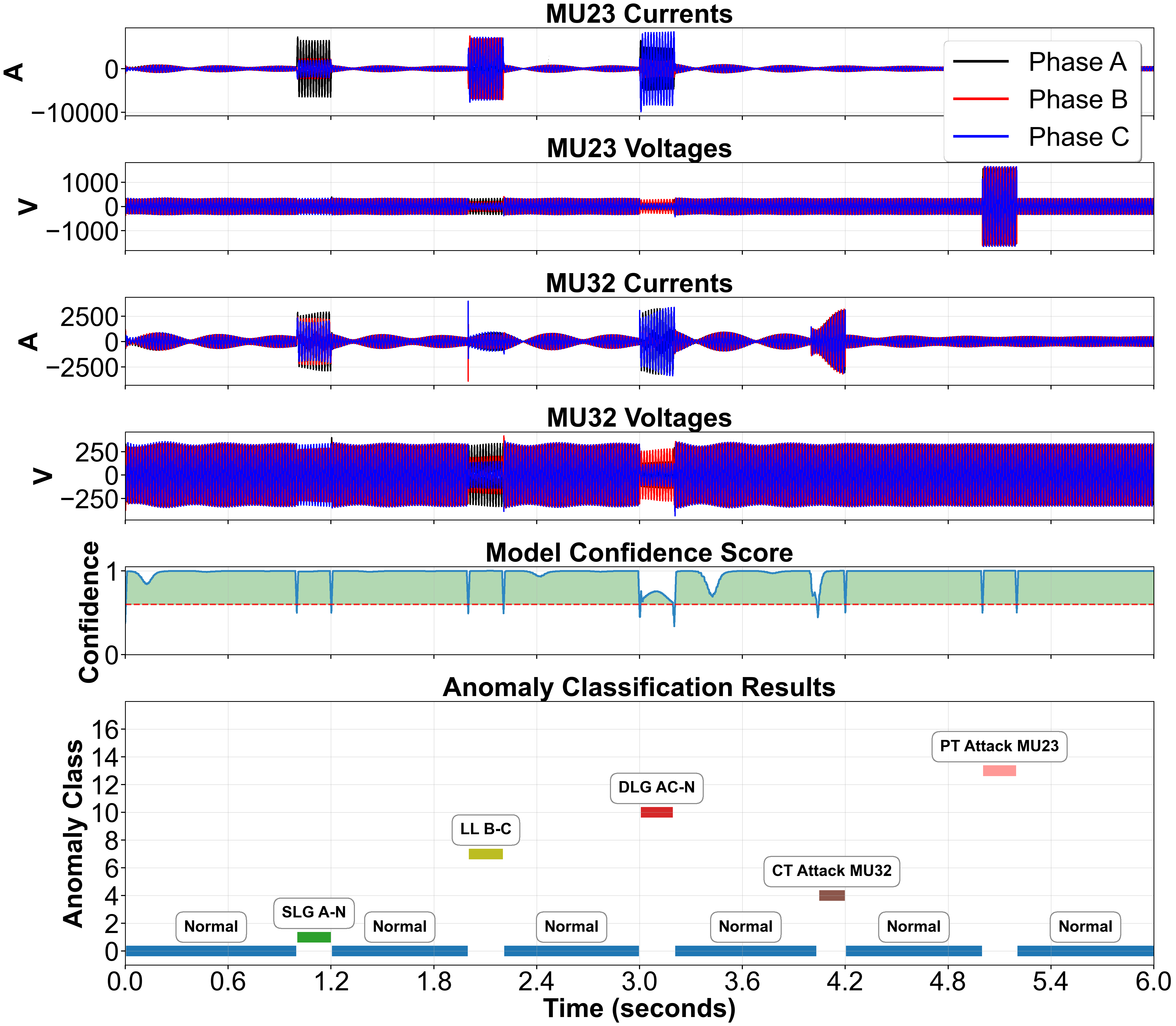}
\caption{MLP neural network response to five testing anomalies showing waveforms and confidence scores}
\label{fig:mlp_response}
\end{figure}

Fig.~\ref{fig:model_comparison} provides a side-by-side view of model performance across all test cases. Detection capability varied widely: some models failed to detect certain anomalies entirely, while the best performers successfully identified all five events. The DLG AC-N fault (shown in red) consistently emerged as the most difficult event to classify. Fig.~\ref{fig:mlp_response} further illustrates the temporal confidence profile of the best-performing MLP model across the complete 6-second evaluation window.

\begin{figure}[!t]
\centering
\includegraphics[width=\columnwidth]{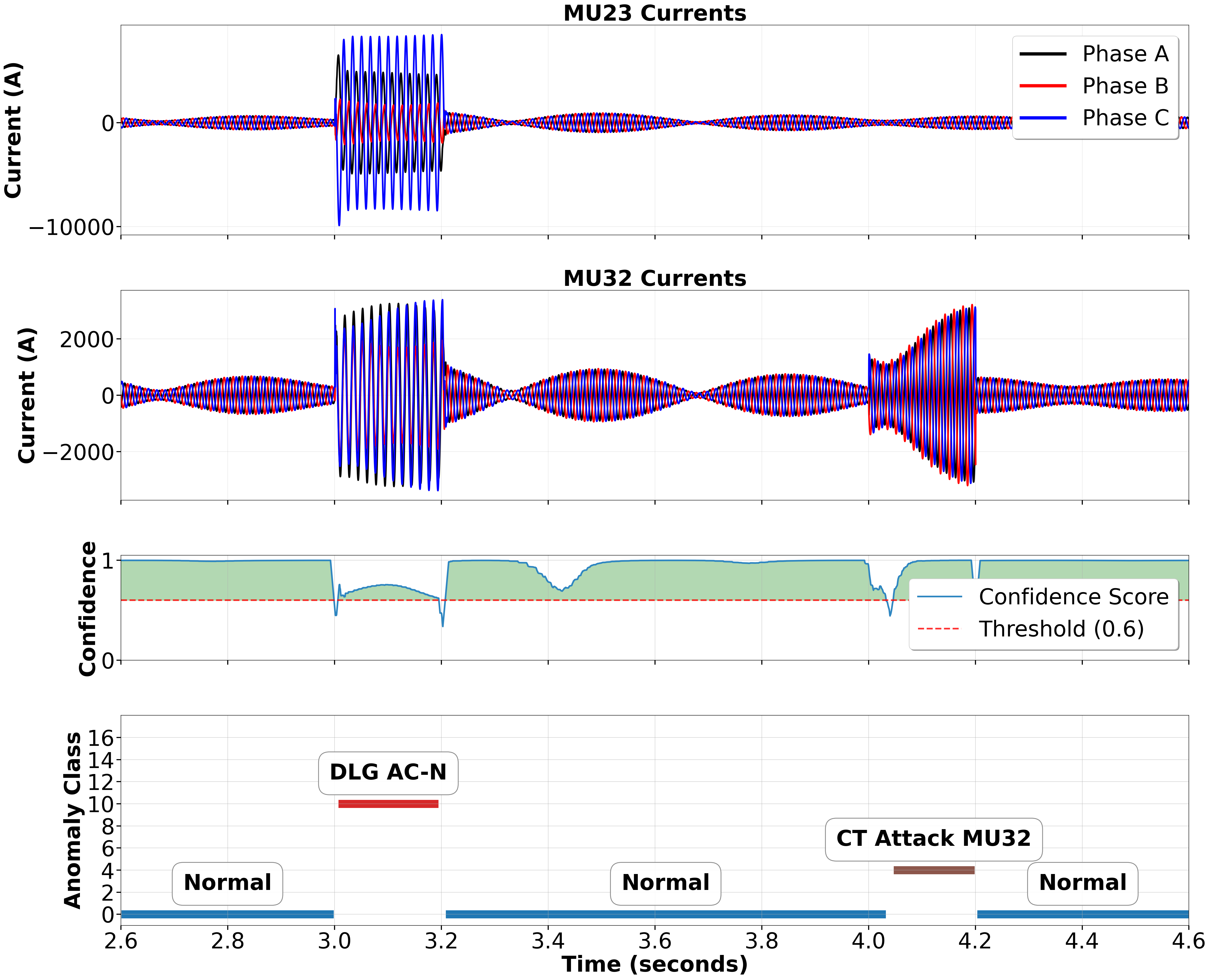}
\caption{Zoomed-in view of the MLP model response to an LLG fault and CT attack, demonstrating correct classification of inter-event transients}
\label{fig:mlp_zoom}
\end{figure}

To better understand classification dynamics, Fig.~\ref{fig:mlp_zoom} zooms in on the 2.6--4.6~s interval. This captures two consecutive but distinct events: the challenging DLG AC-N fault, followed by a CT ratio attack. The model correctly identified the steady-state portion of both anomalies. Notably, it classified the transient recovery period between these events as ``Normal Operation.'' This is a desirable and realistic behavior, as it reflects the model's ability to differentiate between genuine faults and system recovery transients—an essential capability for reducing false positives in protection systems. This robustness can be attributed to the high-fidelity training dataset, which captures IBRs dynamics, damping effects, and fault current limitations.

\begin{figure}[!t]
\centering
\includegraphics[width=\columnwidth]{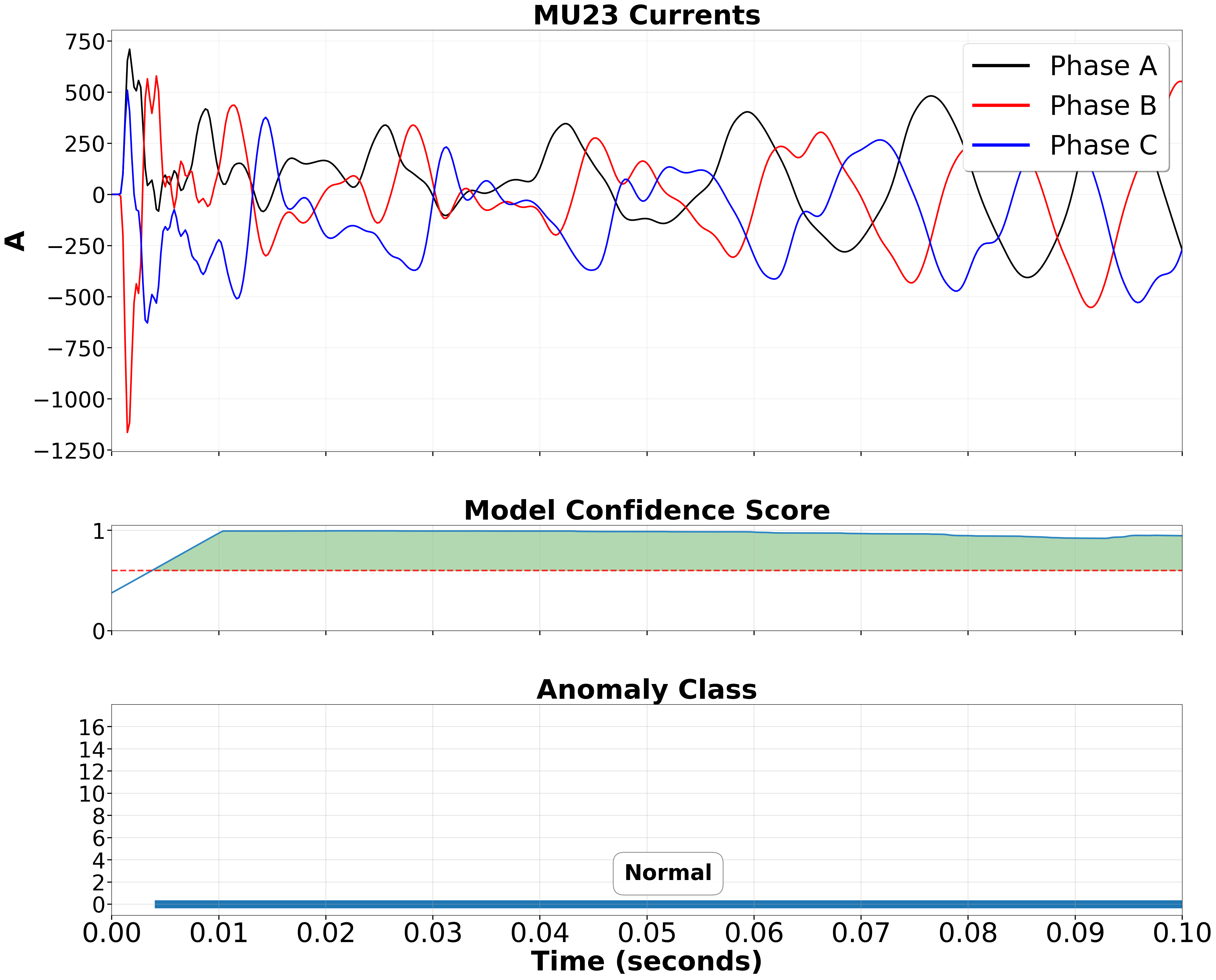}
\caption{MLP model response during initial system energization, correctly classifying high inrush currents as normal operation}
\label{fig:mlp_initial}
\end{figure}

The resilience of the MLP model to non-fault operational disturbances was also evaluated. Fig.~\ref{fig:mlp_initial} shows the initial 0.15~s of the simulation, which includes severe energization transients such as high inrush currents from capacitor bank switching and sub-cycle oscillations due to LC resonance and IBR converter behavior. Despite these conditions—often visually similar to fault signatures—the MLP maintained a ``Normal Operation'' classification throughout. This demonstrates strong discrimination capability between benign start-up disturbances and actual anomalies.

In terms of timing performance, the MLP achieved the best results, classifying events in under 15 ms. However, its average inference latency, including computation, was about 60 ms—exceeding the 50 ms relay requirement and limiting immediate protection deployment \cite{meliopoulos2023dynamic}.

\section{Discussion}
The results highlight a significant divergence between static and streaming evaluations. While most models reached near-perfect accuracies in Phase~1, these results were inflated by class balance and the absence of transient noise. In contrast, Phase~2 streaming tests revealed substantial drops in coverage for ensembles such as Random Forest and Extra Trees, which often abstained from classification despite maintaining accuracy on the cases they did detect. Neural networks, particularly MLPs, proved more resilient, sustaining high coverage and dependable classification under unseen events.

Another key observation is that smoothing and confidence thresholding were critical to stabilize streaming decisions. These mechanisms reduced spurious misclassifications during transient recovery periods and helped distinguish genuine disturbances from benign oscillations, such as capacitor inrush and converter start-up behavior. Nevertheless, even with these improvements, timing analysis revealed that MLPs, although the most dependable models, still exhibited an average inference latency of around 60~ms—above the 50~ms relay requirement. This indicates that while ML-based classification is feasible, immediate application to primary protection is constrained by computational overhead.

Both traditional and ML-based approaches have notable limitations. Conventional dynamic state estimation methods require high-fidelity system models to remain effective, and although they can distinguish between cyberattacks and physical faults, they often provide only coarse-level classifications rather than fine-grained distinctions \cite{abukhousa2025centralized}. Conversely, ML methods can deliver granular, data-driven classifications, but they rely heavily on training data. Any change in system configuration, topology, or operating conditions typically necessitates retraining to preserve reliability. Thus, while ML enhances adaptability and detection speed, ensuring dependable deployment will require a careful balance between physics-based rigor and data-driven flexibility, supported by diverse and evolving datasets.

\section{Conclusion}

This study benchmarked twelve machine learning models for disturbance classification on a high-fidelity, MU-based digital substation dataset, revealing a persistent generalization gap where near-perfect offline accuracies (up to 99.88\%) contrasted with reduced streaming coverage (10--48\% for ensembles). MLPs sustained the highest classification coverage (\(>\)98\%) and stable behavior with cycle-based smoothing and thresholding, whereas ensembles abstained more often, preserving accuracy but reducing coverage. These findings show offline accuracy alone is not a reliable proxy for field readiness, emphasizing the need for streaming-aware evaluation with coverage and abstention metrics in IBR-rich systems. The open release of dataset and code provides a reproducible baseline for future comparisons.

To enhance field deployability, future work will prioritize reducing inference latency of deep learning models to meet sub-cycle relay requirements (\(<\)50~ms), alongside validation on physical testbeds. Additional efforts will explore leakage-free multi-zone datasets for joint classification and localization, robustness under topology changes, and continual-learning strategies to ensure dependable performance over time.

\bibliographystyle{IEEEtran}

\bibliography{references}

@article{sridhar2012cyber,
  title={Cyber-physical system security for the electric power grid},
  author={Sridhar, S. and Hahn, A. and Govindarasu, M.},
  journal={Proc. IEEE},
  volume={100},
  number={1},
  pages={210--224},
  year={2012},
  publisher={IEEE}
}

@article{reno2021influence,
  author={Reno, Matthew J. and Brahma, Sukumar and Bidram, Ali and Ropp, Michael E.},
  journal={IEEE Power and Energy Magazine}, 
  title={Influence of Inverter-Based Resources on Microgrid Protection: Part 1: Microgrids in Radial Distribution Systems}, 
  year={2021},
  volume={19},
  number={3},
  pages={36-46},
  doi={10.1109/MPE.2021.3057951}}

@article{reno2021challenges,
  title={Challenges in Microgrid Protection},
  author={Reno, M. J. and Brahma, S. and Bidram, A. and Ropp, M. E.},
  journal={IEEE Power and Energy Magazine},
  volume={19},
  number={2},
  pages={34--43},
  year={2021},
  publisher={IEEE}
}

@article{manias2024trends,
  title   = {Trends in Smart Grid Cyber-Physical Security: Components, Threats, and Solutions},
  author  = {Manias, Dimitrios M. and Saber, Ahmed M. and Radaideh, Mohamad I. and Gaber, Ahmed T. and Maniatakos, Michail and Zeineldin, Hany and Svetinovic, Davor and El-Saadany, Ehab F.},
  journal = {IEEE Access},
  volume  = {12},
  pages   = {161329--161356},
  year    = {2024},
  doi     = {10.1109/ACCESS.2024.3477714}
}

@article{sahani2023machine,
  title   = {Machine Learning--Based Intrusion Detection for Smart Grid Computing: A Survey},
  author  = {Sahani, Nishita and Zhu, Ronghua and Cho, Jin-Hee and Liu, Chen-Ching},
  journal = {ACM Transactions on Cyber-Physical Systems},
  volume  = {7},
  number  = {2},
  articleno= {11},
  year    = {2023},
  doi     = {10.1145/3578366}
}

@article{ashok2017cyber,
  title   = {Cyber Physical Attack-Resilient Wide-Area Monitoring, Protection, and Control for the Power Grid},
  author  = {Ashok, Aditya and Govindarasu, Manimaran and Wang, Jianhui},
  journal = {Proceedings of the IEEE},
  volume  = {105},
  number  = {7},
  pages   = {1389--1407},
  year    = {2017},
  doi     = {10.1109/JPROC.2017.2686391}
}

@article{zonouz2012scpse,

  author={Zonouz, Saman and Rogers, Katherine M. and Berthier, Robin and Bobba, Rakesh B. and Sanders, William H. and Overbye, Thomas J.},
  journal={IEEE Transactions on Smart Grid}, 
  title={SCPSE: Security-Oriented Cyber-Physical State Estimation for Power Grid Critical Infrastructures}, 
  year={2012},
  volume={3},
  number={4},
  pages={1790-1799},
  doi={10.1109/TSG.2012.2217762}
}

@article{li2023identification,
  title={Identification and classification for multiple cyber attacks in power grids based on deep capsule convolutional neural network},
  author={Li, Y. and others},
  journal={Eng. Appl. Artif. Intell.},
  volume={124},
  pages={106572},
  year={2023},
  publisher={Elsevier}
}

@inproceedings{alkasassbeh2021ensemble,
  title={Ensemble learning methods for anomaly intrusion detection system in smart grid},
  author={Alkasassbeh, M. and Alauthman, M. and Alweshah, M.},
  booktitle={Proc. IEEE Jordan Int. Joint Conf. Elect. Eng. Inf. Technol.},
  pages={1--6},
  year={2021},
  organization={IEEE}
}

@article{al2023ensemble,
  title={Ensemble voting-based anomaly detection for a smart grid communication infrastructure},
  author={Al-Abassi, A. and others},
  journal={Intell. Autom. Soft Comput.},
  volume={36},
  number={3},
  pages={3257--3278},
  year={2023}
}

@inproceedings{johnson2014machine,
  author={Borges Hink, Raymond C. and Beaver, Justin M. and Buckner, Mark A. and Morris, Tommy and Adhikari, Uttam and Pan, Shengyi},
  booktitle={2014 7th International Symposium on Resilient Control Systems (ISRCS)}, 
  title={Machine learning for power system disturbance and cyber-attack discrimination}, 
  year={2014},
  volume={},
  number={},
  pages={1-8},
  doi={10.1109/ISRCS.2014.6900095}}

@article{gholami2022active,
  title={Active distribution system co-ordinated control method via artificial intelligence},
  author={Gholami, A. and Aligholian, A. H.},
  journal={arXiv preprint arXiv:2207.14642},
  year={2022}
}

@article{meliopoulos2023dynamic,
  title={Dynamic Estimation-Based Protection and Hidden Failure Detection and Identification: Inverter-Dominated Power Systems},
  author={Meliopoulos, S. and Cokkinides, G. J. and Myrda, P. and Farantatos, E. and Elmoudi, R. and Fardanesh, B. and Stefopoulos, G. and Black, C. and Panciatici, P.},
  journal={IEEE Power \& Energy Mag.},
  volume={21},
  number={1},
  pages={59--71},
  year={2023},
  publisher={IEEE}
}

@article{chencaisakis2024dl,
  author={Chen, Z. and Cai, S. and Meliopoulos, A. P. Sakis},
  booktitle={2024 IEEE Power & Energy Society Innovative Smart Grid Technologies Conference (ISGT)},
  title={A Real-time Deep Learning-based Fault Diagnosis Framework in Power Distribution System with PVs},
  year={2024},
  volume={},
  number={},
  pages={1-5},
  doi={10.1109/ISGT59692.2024.10454179},
  ISSN={},
  month={Feb.},
}

@inproceedings{alsaeed2025anomaly,
  author={Alsaeed, Fahad and Abukhousa, Emad and Afroz, Syed Sohail Feroz Syed and Qwbaiban, Abdulaziz and Sakis Meliopoulos, A.P.},
booktitle={2025 IEEE International Conference on Cyber Security and Resilience (CSR)}, 
title={Anomaly Identification in Power Systems Using Dynamic State Estimation and Deep Learning}, 
year={2025},
volume={},
number={},
pages={530-536},
doi={10.1109/CSR64739.2025.11130018}}

@manual{winigs2025,
  title        = {{WinIGS} Integrated Grounding System Analysis for Windows -- Version 8.1.5},
  organization = {Advanced Grounding Concepts (AGC)},
  address      = {Alpharetta, GA, USA},
  year         = {2025},
  month        = {May},
  note         = {Proprietary power system analysis and grounding simulation software. Latest version accessed May 30, 2025},
  url          = {https://ap-concepts.com/}
}

@techreport{pserc2018t59g,
  author      = {A. P. Sakis Meliopoulos and George Cokkinides and Jiahao Xie and Yuan Kong},
  title       = {RTE DSE Protection Demonstration},
  institution = {Power Systems Engineering Research Center (PSERC)},
  number      = {PSERC Publication 18-09, Final Project Report T-59G},
  address     = {Tempe, AZ},
  year        = {2018},
  month       = {September},
  url         = {https://documents.pserc.wisc.edu/documents/publications/reports/2019_reports/T_59G_Final_Report__2_.pdf},
  note        = {Includes modeling and event simulation using WinIGS-T}
}

@article{abukhousa2025centralized,
  title={Centralized Dynamic State Estimation Algorithm for Detecting and Distinguishing Faults and Cyber Attacks in Power Systems},
  author={Abukhousa, Emadeldin A. and Syed Afroz, Syed Sohail Feroz and Alsaeed, Fahad and Qwbaiban, Abdulaziz and Meliopoulos, A.P. Sakis},
  journal={arXiv preprint arXiv:2508.02102},
  year={2025},
  doi={10.48550/arXiv.2508.02102}
}

@misc{T4Tech,
	author       = {Emad Abukhousa},
	title        = {T4Tech: Open Dataset and Code for Power System Fault and Cyber-Attack Classification},
	year         = {2025},
	howpublished = {\url{https://github.com/Emadeddin/T4Tech}},
	note         = {Accessed: August 28, 2025}
}

\end{document}